\documentclass[conference]{IEEEtran}

\IEEEoverridecommandlockouts
\usepackage{cite}
\usepackage{amsmath,amssymb,amsfonts}
\usepackage{algorithmic}
\usepackage{graphicx}
\usepackage{textcomp}
\usepackage{hyperref}
\hypersetup{colorlinks=true, unicode=true, linkcolor=[rgb]{0.10,0.05,0.67}, citecolor=[rgb]{0.10,0.05,0.67}, filecolor=[rgb]{0.10,0.05,0.67}, urlcolor=[rgb]{0.10,0.05,0.67}}
\usepackage{dblfloatfix}
\usepackage{cases}
\usepackage{nomencl}
\usepackage{flushend}
\usepackage{etoolbox}
\usepackage{dirtytalk}
\usepackage{tabulary}
\usepackage{colortbl}
\usepackage{soul}
\usepackage[normalem]{ulem}
\usepackage{amsmath}
\newcommand{\stkout}[1]{\ifmmode\text{\sout{\ensuremath{#1}}}\else\sout{#1}\fi}

\usepackage[symbol]{footmisc}
 \newcommand\footnoteref[1]{\protected@xdef\@thefnmark{\ref{#1}}\@footnotemark}

\usepackage{draftwatermark}
\SetWatermarkText{Accepted for IEEE PES GM 2023}
\SetWatermarkScale{0.3}
\SetWatermarkColor[gray]{0.9}

\renewcommand\nomgroup[1]{%
  \item[\bfseries 
  \ifstrequal{#1}{A}{\normalfont{\textbf{\textit{Sets (indices)}}}}{%
  \ifstrequal{#1}{B}{\normalfont{\textbf{\textit{Variables}}}}{%
  \ifstrequal{#1}{C}{\normalfont{\textbf{\textit{Parameters}}}{}}}}%
]}

\newcommand{\RomanNumeralCaps}[1]
    {\MakeUppercase{\romannumeral #1}}
    
\renewcommand{\figurename}{Fig. }

\usepackage{xcolor}

\def\BibTeX{{\rm B\kern-.05em{\sc i\kern-.025em b}\kern-.08em
    T\kern-.1667em\lower.7ex\hbox{E}\kern-.125emX}}
\begin{document}

\title{Impact of Dynamic Tariffs for Smart EV Charging on LV Distribution Network Operation\\

\thanks{}
}

\author{\IEEEauthorblockN{Flore Verbist, Nanda Kishor Panda, Pedro P. Vergara, Peter Palensky}
\IEEEauthorblockA{Electrical Engineering, Mathematics and Computer Science,
Delft University of Technology, Delft, The Netherlands \\
Email: f.verbist@student.tudelft.nl, \{n.k.panda, p.p.vergarabarrios, p.palensky\}@tudelft.nl}
 }

\maketitle

\begin{abstract}
With a growing share of electric vehicles (EVs) in our distribution grids, the need for smart charging becomes indispensable to minimise grid reinforcement. To circumvent the associated capacity limitations, this paper evaluates the effectiveness of different levels of network constraints and different dynamic tariffs, including a dynamic network tariff. A detailed optimisation model is first developed for public charging electric vehicles in a representative Dutch low voltage (LV) distribution network, susceptible to congestion and voltage problems by 2050 without smart charging of EVs. Later, a detailed reflection is made to assess the influence of the modelled features on the distribution system operator (DSO), charge point operator (CPO) costs, and the EVs' final state-of-charge (SOC) for both mono- (V1G) and bi-directional (V2G) charging. Results show that the dynamic network tariff outperforms other flat tariffs by increasing valley-filling. Consequently, compared to regular day-ahead pricing, a {significant} reduction in the frequency of congestion in the lines is achieved. In addition, V2G ensures the joint optimum for different stakeholders causing adequate EV user satisfaction, decreased CPO costs  compared to conventional charging and fewer violations of grid constraints for the DSOs.

\end{abstract}


\vspace{-0.23 cm}
\makenomenclature
\vspace{-0.3 cm} 
\nomenclature[A]{$\Omega_{\mathrm{T}}(t)$}{ $ \text{Set of horizon time steps; }  t \in \mathbb{N}$.}
\nomenclature[A]{$\Omega_{\mathrm{C}}(c)$}{$ \text {Set of charging points; } c \in  \mathbb{N} $.}
\nomenclature[A]{$\Omega_{\mathrm{B} }(b)$}{$ \text {Set of nodes; } b \in \mathbb{N}_{0}. $}
\nomenclature[A]{$\Omega_{\mathrm{L}}(l)$}{$ \text {Set of lines; } l \in  \mathbb{N}_{0} \times \mathbb{N}_{0}:  l =(i,j)  \in \Omega_{\mathrm{B}} \times\Omega_{\mathrm{B}}  \land i \neq j \land i,j = \text{binding nodes} $.}
\nomenclature[A]{$\Omega_{\mathrm{P}}(p)$}{ \text {Set of low-, medium- and high-level network tariff} components; $ p = \{ \text {ll, ml, hl} \}. $}

\nomenclature[B]{$\mathrm{{P}}_{t,c}$}{Active (dis)charging power for EV at charging point $c$ at time $t$.}

\nomenclature[B]{$\mathrm{P}_{t}^{ \text{tf}}$}{Aggregated transformer power without power losses.}
\nomenclature[B]{$\mathrm{P}_{t}^{ \text{tf}^{*}}$}{Aggregated transformer power using linear power flow equations.}
\nomenclature[C]{$\overline{\mathrm{P}}^{ \text{tf}}$}{Rated transformer power (400 kVA).}
\nomenclature[B]{$\mathrm{P}_{t}^{{p}}$}{Low, medium and high tariff charging power at transformer; $\forall p \in \Omega_{\mathrm{P}}$.}
\nomenclature[B]{$\mathrm{P}_{t}^{ \text{Dis}}$}{Net discharge power at transformer;  $\mathrm{P}_{t}^{ \text{Dis}} \in \mathbb{R}^{-}$.}
\nomenclature[B]{$\mathcal{V}_{t,c}$}{SOC (\%) of EV at charging point c connected at time $t$.}
\nomenclature[B]{$\alpha_{c}$}{Fraction of the {maximum} SOC at departure; $\alpha_c \in [0,1]$.}
\nomenclature[B]{$\mathbf{I}_{t,l, \phi}$}{Phase current at line $l$ for phase $\phi$.}
\nomenclature[B]{$\mathbf{V}_{t,b,\phi}$}{Phase Voltage at node $b$ for phase $\phi$.}

\nomenclature[C]{$\Delta \mathrm{t}$}{Duration of one time period (0.25 h).}
\nomenclature[C]{$\mathrm{c}_{t}^{ \text {DA}}$}{Day-ahead price at time $t$.}
\nomenclature[C]{$\mathrm{c}^{{p}}$}{Low, medium and high network tariff cost of charging;  $\forall p \in \Omega_{p}$.}

\nomenclature[C]{$t^{\text{dep}}_{c}$}{Departure time of EV at charging point $c$.}
\nomenclature[C]{$\overline{\mathrm{P}}_{c}$}{Maximum rated power of EV at charging point $c$.}

\nomenclature[C]{$\mathcal{V}^{\text{init}}_c$}{Initial SOC percentage of EV at charging point $c$.}

\nomenclature[C]{$\overline{\mathcal{V}}_c$}{Maximum SOC percentage of EV at charging point $c$.}
\nomenclature[C]{$\underline{\mathcal{V}}_c$}{Minimum SOC percentage of EV at charging point $c$.}
\nomenclature[C]{$\overline{\mathrm{E}}_{c}$}{Maximum battery capacity of EV at charging point $c$.}
\nomenclature[C]{$\overline{\mathrm{P}}_{t}^{ {p}}$}{Available low, medium and high cost charging power at time $t$; $\overline{\mathrm{P}}_{t}^{{p}} \in \mathbb{R^{+}}, \: \forall \: p \in  \Omega_p.$ }

\nomenclature[C]{$\mathrm{\mu}_{c}^{\text{V2G}}$}{Binary parameter indicating bi- (1) or mono-directional (0) charging capability.}
\nomenclature[C]{$\mathrm{M}$}{Large positive (big M) constant for variable $\alpha_{c}$.}
\printnomenclature
\section{Introduction}
\IEEEPARstart{D}{riven} by the goal to reach carbon neutrality by 2050, EVs are being scaled up to become an integral part of the streetscape in Europe. However, the massive deployment of EVs will stress the already congested power networks, making them more vulnerable to failures. This effect is more visible on low voltage (LV) distribution networks, which traditionally are not built to withstand such exponential growth in loading levels~\cite{Verbist2022InterplayThesis}. As EVs are parked for 95\% of the time, they enable flexibility to provide ancillary services to power networks~\cite{IRENA2019InnovationVehicles}, potentially solving the issues caused by them in the first place. Especially off-peak charging is a requirement for the safe operation of our existing LV networks. The EV's flexibility can be enabled with smart charging, often using cost-based objectives~\cite{Hennig2020CapacityManagement, Limmer2019DynamicReview, Panda2021APark}.

Capacity subscription tariffs could help to spread consumer loading more, as analysed in~\cite{Hennig2020CapacityManagement}, but do not reflect changing grid conditions and would suppress the power of EV flexibility provided to the grid. The same holds for time-of-use (TOU) tariffs that apply two or three different but static price levels~\cite{Nimalsiri2021CoordinatedBenefit, Esmaili2015Multi-objectiveNetworks}. Real-time pricing with dynamic day-ahead prices could better represent the grid conditions, such as a change in the renewable energy share, and unlock more EV flexibility. However, day-ahead price-based charging can shift load peaks to other times of the day with low prices instead of reducing them~\cite{Limmer2019DynamicReview}. Therefore, tariffs that capture the varying load at the transformer level can provide better peak shaving, as shown in \cite{ DeA.Bitencourt2017OptimalPricing}. The shortcoming of using dynamic costs at the transformer level is that congestion and voltage problems can still occur locally downstream of the transformer. This is an important aspect to look at for LV networks, which are more prone to voltage problems and congestion due to the rising use of distributed renewable energy sources such as photovoltaic (PV) systems. This problem can be catered to by including network constraints and the associated power flow equations inside the optimisation model used to dispatch EVs. In the current literature, various equation-based power flow models vary in computational efficiencies and accuracy ~\cite{Giraldo2021ANetworks, Nimalsiri2021CoordinatedBenefit, deHoog2015OptimalAccount, Franco2015ASystems}. For the operational dispatch of EVs, smart charging EV fleets to minimise costs for the charge point operators (CPOs) can conflict with the grid's reliable operation. Authors in \cite{Panda2021APark} show how a multi-objective optimisation framework can quantify cost saving along with minimising the grid's in-feed power for a multi-energy system from a planning perspective. However, such an approach is unsuitable for the operational dispatch of EVs as solving multi-objective optimisation problems for a whole year using a lexicographic approach takes longer than EVs' operational dispatch time.

{ The aim of this paper is to test the EVs' charging flexibility in relation to different levels of grid constraints and a novel dynamic tariff, recently introduced in a pilot project\footnote[3]{Received as a personal communication from Nico Brinkel, Utrecht University, The Netherlands. The dynamic network tariff implemented in the paper is currently being used in \href{https://ssc-fleet.nl/}{FLEET} project as a smart charging pilot.}. For that reason, an operational dispatch model including receding horizon optimisation (RHO) to allow optimal charging of EVs is developed. The model is validated with power flow studies of a real Dutch LV grid including future-proof data and charging technology such as vehicle-to-grid (V2G). The contribution of this paper is twofold. First, compared to the discussed literature, the tariff as well as the modelled grid constraints should allow charging EVs in accordance with expected changes in distributed power generation and consumption.
Second, this study distinguishes itself from others by catering for the need of all the stakeholders involved, namely: CPO, DSO, and end users. The analyses provide useful insights into the available flexibility of EV charging in relation to grid reinforcement needed in LV grids with 100\% EV penetration.}

\vspace{-0.23 cm}
\section{Optimisation Framework}
\label{sec:optimisation}
An RHO model with a planning horizon of 24 hours and 15 minutes intervals is formulated. The chosen time frame allows us to better plan EV dispatch and unlocks enough EV flexibility, especially during long connection times. For the entire planning horizon, information about the expected non-EV loads at each of the LV nodes and day-ahead prices is taken as input. The model needs to be initialised with the initial state-of-charge (SOC), maximum charging power, V2G information and EVs' battery capacity. After the initialisation of SOC in the first step, the rest of the stages use the information about SOC calculated in the previous stage. However, information about the arrival of new vehicles and the estimated departure time of the vehicles is received as an external input in a rolling fashion as the EVs arrive at the charging stations. 

\begin{equation}
\hspace*{-1cm} 
 \! { \min_{P_{t,c}} } \! \sum_{t\in \Omega_{\mathrm{T}}} \! {\! \left(
    \overbrace{c_t^{DA} \! \sum_{\! \mathrm{c}\in \Omega_\mathrm{C}}{\! \mathrm{P}_{t,c}}}^{\text{\RomanNumeralCaps 1} \! } \!+ \! \underbrace{\sum_{p\in \Omega_\mathrm{P}}{ \! \mathrm{c}^p \mathrm{P}_t^p}}_{\text{\RomanNumeralCaps 2} } \! + \! \overbrace{\mathrm{P}_t^{\text{tf}^{*}}}^{\text{\RomanNumeralCaps 3} } \! - \!
    \underbrace{\! \sum_{c\in \Omega_\mathrm{C}}{\! \mathrm{M}\alpha_c}}_{\text{\RomanNumeralCaps 4} \!}\right) \! \Delta \mathrm{t}}  \! \label{eq:OBJf}
\end{equation}
subject to:
 \begin{alignat}{2}
\underline{\mathcal{V}}_c \leq \mathcal{V}_{t,c} \leq \alpha_c \overline{\mathcal{V}}_c  \quad\quad 
\quad ;\forall \:\: t \: \in \{t_{c}^{\text{dep}}\},\: c \in \Omega_\mathrm{C} &
    \label{eq:SOC_limits} \\
    \mathcal{V}_{t,c}=\begin{cases}
                    \mathcal{V}_c^{\text{init}} &;\forall \:\: t \: \in \{1\},\: c \in \Omega_\mathrm{C}  \\[2 pt]
                    \mathcal{V}_{t-1,c} + \Delta \mathrm{t} \frac{\mathrm{P}_{t-1,c}}{\overline{\text{E}_{c}}} &  ;\forall \:\: t \: \in \Omega_{\mathrm{T}}\text{-}\{1\},\: c \in \Omega_\mathrm{C} 
          \end{cases} 
        \label{eq:SOC_update} 
\end{alignat}
\begin{alignat}{2}
   - \mu_{c}^{\text{V2G}} \overline{\mathrm{P}_{c}} \leq \mathrm{P_{t,c}}\leq \overline{\mathrm{P}_{c}} & \quad\quad;  \forall \:  t \in \Omega_{\mathrm{T}}
  \label{eq:P_limits} \\
\mathrm{P}_{t}^{p} \leq \overline{\mathrm{P}}_{t}^{p}                                 & \quad\quad; \forall \:\: t \: \in \Omega_{\mathrm{T}},\: p \in \Omega_p  
\label{eq:Stacked_limit}\\
\sum_{c\in \Omega_\mathrm{C}}\mathrm{P}_{t,c} = \sum_{p\in \Omega_{\mathrm{P}}}\mathrm{P}_t^p + \mathrm{P}_t^{\text{Dis}} & \quad \quad ;\forall \:\: t \: \in \Omega_{\mathrm{T}}
\label{eq:Stacked_eq} \\
\mathrm{P}_t^{\text{tf}} \leq \overline{\mathrm{P}}^{\text{tf}} & \quad\quad ;  \forall \:\: t \:\in \Omega_{\mathrm{T}} \label{eq:trf_limit} \\
\left\|\mathbf{I}_{t, l}\right\|_{2} \leq \overline{\mathrm{I}}_{l} & \quad\quad;  \forall \:\: t \:\in \Omega_{\mathrm{T}},  l  \:  \in \Omega_{\mathrm{L}} \label{eq:current} \\
\underline{\mathrm{V}} \leq\left\|\mathbf{V}_{t,b}\right\|_{2} \leq \overline{\mathrm{V}} &  \quad\quad ; \forall \:\: t \: \in \Omega_{\mathrm{T}},  b \:  \in \Omega_{\mathrm{B}}
\label{eq:voltage} \\
{P_{t, b}=\mathfrak{R e}\left\{\mathbf{V}_{t, b}^{\top} \mathbf{I}_{t, b}^{\mathbf{}^*}\right\},}  &  \quad {Q_{t, b}=\mathfrak{I} \mathfrak{m}\left\{\mathbf{V}_{t, b}^{\top} \mathbf{I}_{t, b}^{\mathbf{}^*}\right\}}
\label{eq:power}
\end{alignat}

For all time steps of the planning horizon, the individual charging power at each charging point is optimised for the objective function given in \eqref{eq:OBJf}. The proposed objective function is minimised so that the interests of all involved stakeholders are catered for. It implies minimising the CPO costs (components \RomanNumeralCaps 1 \& \RomanNumeralCaps 2), maximising the SOC of the EVs (\RomanNumeralCaps 4) and minimising power losses, congestion and voltage issues to comply with DSO interests using dynamic, stacked tariffs (\RomanNumeralCaps 2) and power flow modelling (\RomanNumeralCaps 3).

\begin{figure}[bt]
\centering
\includegraphics[width=\linewidth]{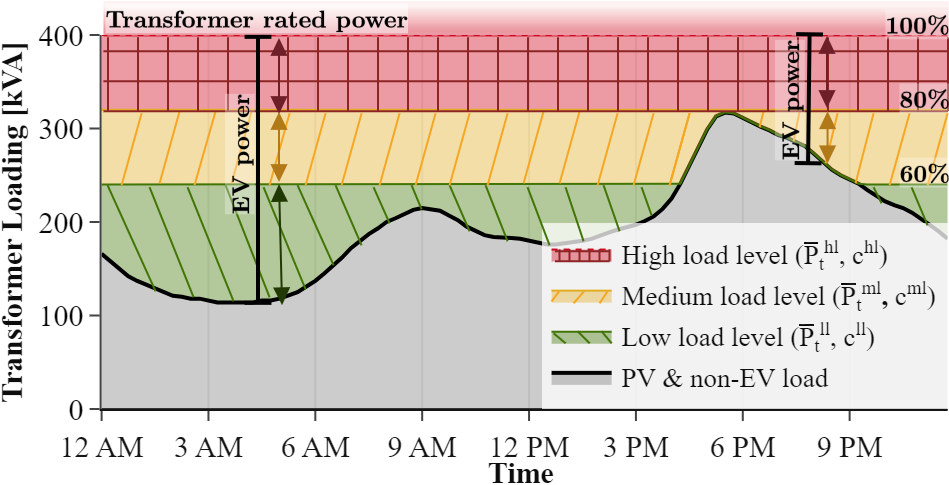}
\caption{Schematic illustrating the dynamic network component of the stacked tariff.}
\label{stacked_tariff}
\end{figure}

The new stacked dynamic tariff, managed by the DSO and charged to the CPO, consists of two dynamic components: the day-ahead electricity costs (\RomanNumeralCaps 1) and the dynamic network cost (\RomanNumeralCaps 2). The latter is explained by \figurename{\ref{stacked_tariff}} and constraints in \eqref{eq:Stacked_eq}, \eqref{eq:Stacked_limit}. Three loading levels are defined, each with a certain maximum time-varying capacity ($\mathrm{\overline{P}}_{t}^{p}$) and a fixed cost ($\mathrm{c}^{p}$). The following relation holds: ${\mathrm{c}}^{\text{ll}} < {\mathrm{c}}^{\text{ml}} < \mathrm{{c}}^{\text{hl}}$, with ${\mathrm{c}}^{\text{ll}}$ the lowest cost level and ${\mathrm{c}}^{\text{hl}}$ the highest. The corresponding capacity levels are predetermined at 60\%, 80\% and 100\% of the transformer-rated power\footnotemark[3], but the actual volume assigned to them varies depending on the predicted transformer loading excluding EVs, as seen in \figurename\ref{stacked_tariff}. For instance, during 4 - 9 pm (c.f. \figurename\ref{stacked_tariff}), only medium and high network costs can be assigned to EV charging due to high expected transformer loading. Consequently, off-peak charging is stimulated if flexibility is available, leaving the CPO with lower network costs and respecting grid limits more. Constraint \eqref{eq:Stacked_eq} ensures that the sum of the optimised load levels, as bounded by \eqref{eq:Stacked_limit}, equals the sum of the individual charging power levels at each time step $t$. 


Component \RomanNumeralCaps 3 in \eqref{eq:OBJf} tries to reduce the aggregated power losses such that optimal power flow modelling is ensured and grid constraints can be applied in the LV network. These grid constraints are covered by \eqref{eq:current} and \eqref{eq:voltage}, limiting the current in the lines and voltage magnitude in the nodes. The values of $\mathrm{I}_{t,l}$ and $\mathrm{V}_{t,b}$ are obtained by solving the linearised optimal power flow (OPF) presented in~\cite{Giraldo2021ANetworks} {as implemented by~\cite{Verbist2022InterplayThesis}}. {The generic power flow relationship is shown by \eqref{eq:power}}. To restrict loading at the transformer level, as expressed by \eqref{eq:trf_limit}, a more simplified formulation of $\mathrm{P}_{t}^{\text{tf}^{*}}$, denoted as $\mathrm{P}_{t}^{\text{tf}}$, could be obtained as well by adding all loads and subtracting generation, omitting power losses in the lines. That excluded the need for computational-demanding power flow modelling in scenarios that only investigated the effect of a transformer limit.

The last term in \eqref{eq:OBJf} maximises the SOC level of each vehicle. The parameter $\mathrm{M}$ penalises lower $\alpha_c$, maximising the final SOC at departure time. This maximises the final SOC at departure time. This is encompassed by \eqref{eq:SOC_limits}, together with the SOC boundaries. Equation \eqref{eq:SOC_update} ensures that the SOC is updated at each time step till departure time. When no EV is connected to the charging point $c$, power and SOC values are set to zero. The charging power is limited by \eqref{eq:P_limits}.

\section{Case Study}
\begin{figure}[t]
    \centering
    \includegraphics[width=0.9\linewidth]{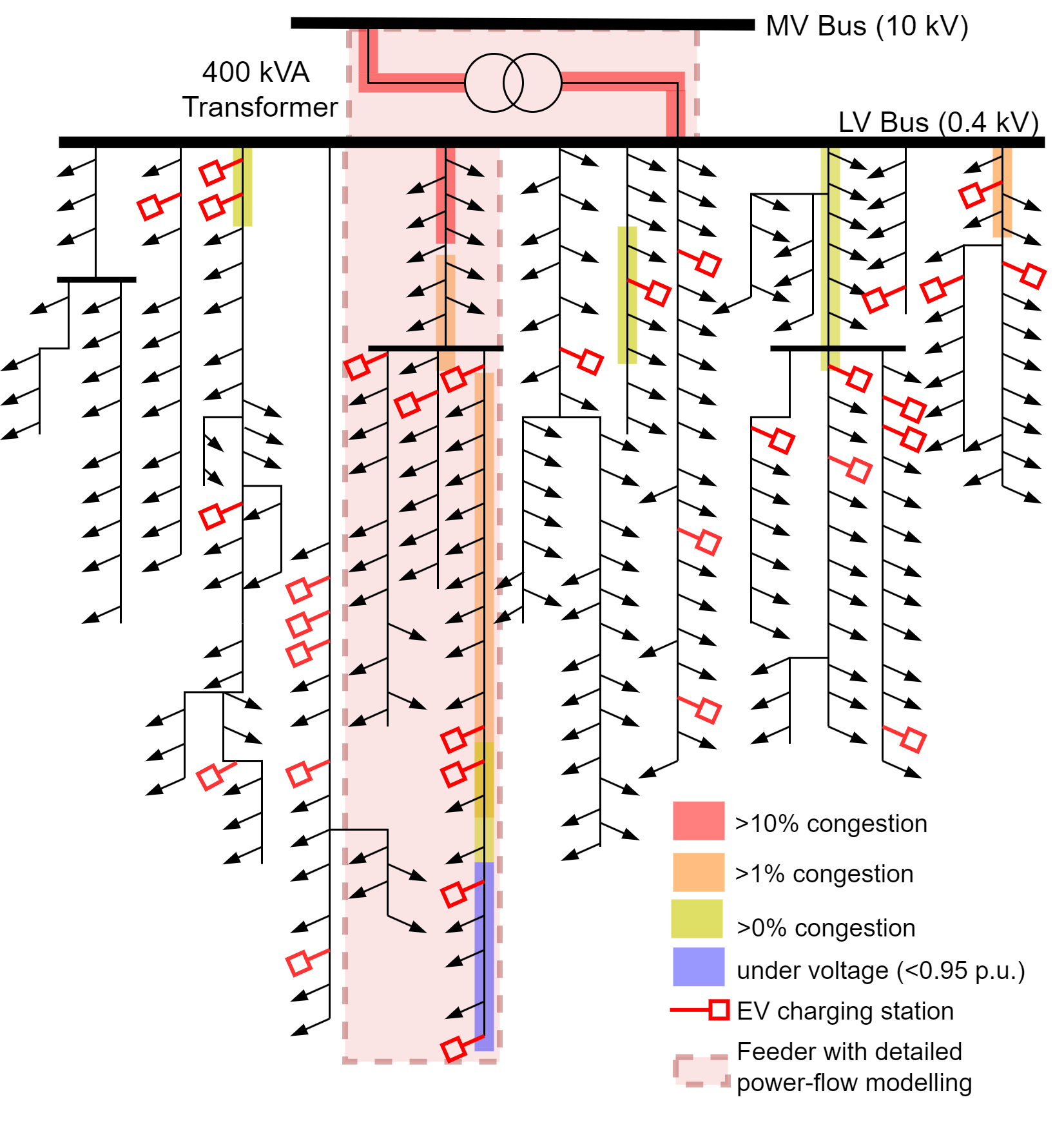}
    \caption{Representative LV grid in the Netherlands used for {the analyses. The red transparent boxes indicate the grid parts covered by detailed power flow modelling in scenario S4 with constraints \eqref{eq:current} and \eqref{eq:voltage} as 95\% of grid issues apply to that feeder.}}
    \label{fig:lv_schematic}
\end{figure}
\begin{table}[b]
\begin{minipage}{\linewidth}
\caption{Scenario Overview }
\begin{center}
\begin{tabular}{ccccc}
\hline
\multicolumn{1}{c}{\textbf{\begin{tabular}[c]{@{}c@{}}Scenario \end{tabular}}} &
  \textbf{Tariff} &
  \textbf{V2G [\%]} &

  \textbf{\begin{tabular}[c]{@{}c@{}}Objective \\ Function\end{tabular}} &
    \textbf{ Constraints} 
  
  \\ \hline\hline
S0 & \textminus       &  0 & \textminus   &     \textminus       \\ 
S1 & Day-ahead & 80  & \RomanNumeralCaps 1, \RomanNumeralCaps 4 & (\ref{eq:SOC_limits} - \ref{eq:P_limits}), \eqref{eq:trf_limit}  \\ 
S2 & Stacked   & 80  & \RomanNumeralCaps 1, \RomanNumeralCaps 2 , \RomanNumeralCaps 4 &  (\ref{eq:SOC_limits} - \ref{eq:trf_limit})  \\ 
    S3 & Stacked & 0 &    \RomanNumeralCaps 1, \RomanNumeralCaps 2, \RomanNumeralCaps 4   &   (\ref{eq:SOC_limits} - \ref{eq:trf_limit})  \\ 
S4 & Stacked  & 80 &  \RomanNumeralCaps 1, \RomanNumeralCaps 2, \RomanNumeralCaps 3, \RomanNumeralCaps 4  &   (\ref{eq:SOC_limits} - \ref{eq:voltage}) \\ \hline
\end{tabular}
\label{tab:scenario}
\end{center}
\end{minipage}
\end{table}
A typical radial urban Dutch LV network (c.f. \figurename{\ref{fig:lv_schematic}}) consisting of 11 feeders and a total of 504 nodes, connected to the MV grid through a 400 kVA (10 kV/ 0.4 kV) transformer is taken as the case study for the analysis presented in this paper. First, uncontrolled charging for a 2050 scenario was investigated. After that, the model presented in Section \ref{sec:optimisation} was applied to the LV grid using the Pyomo optimisation library in Python. Due to the lack of private parking spaces, active involvement of a single CPO and the main objective to maintain the grid quality, this centralised model was preferred over a decentralised model~\cite{Limmer2019DynamicReview, deHoog2013ElectricApproaches}. The smart charging analyses using the optimisation model are applied to the worst-case scenario, one week in the winter of 2050, with 100\% EV penetration (0.4 passenger cars per household). For smart charging, DigSILENT PowerFactory is used to provide network data as an input to the optimisation model and is used to validate the optimisation results on voltage and congestion problems. As one charging pole for every 6 EVs is assumed as the standard requirement, a total of 32 charging stations, each with two charging points, were randomly placed in the distribution grid as shown in \figurename \ref{fig:lv_schematic}.\\
Different scenarios analysed in this paper are presented in Tab. \ref{tab:scenario}, where each scenario differs by the type of tariff used and the degree of network constraints enforced. S0 represents the base case where all the EVs are assumed to charge to their required level as soon as they are connected to a charger, thereby ruling out any possibilities of flexibility or smart charging. Scenarios S1, S2 and S3, represent cases where EV charging is automatically optimised considering only the capacity limitation of the MV/LV transformer \eqref{eq:trf_limit}. For these scenarios, the calculated net power at the transformer neglects any power loss due to the lack of detailed load flow. For S4, equation-based power flow is used to model one of the longest feeders (c.f. \figurename \ref{fig:lv_schematic}) of the LV network based on~\cite{Giraldo2021ANetworks}, which restricts the bus node voltages to $\pm 5\%$ of the nominal voltage. Congestion at the transformer was limited to 100\% of its rated power, excluding power losses, leading to 120\% loading of the transformer and 100\% loading of the transformer line. These values were regarded as acceptable to the DSO. In addition, a sensitivity analysis revealed a linearisation error of 6.5\% compared to the validation in PowerFactory (Newton-Raphson method), such that a correction factor of 0.938 was multiplied with the rated current in \eqref{eq:current}. S1 uses only Day-ahead (DA)\footnote[1]{Prices used from \href{https://www.entsoe.eu/}{ENTSOE}.} prices for the energy imported from the grid in the objective function. It assumes a fixed network tariff (not included in the optimisation), whereas scenarios S2-S4 use a stacked tariff scheme. The stacked tariff scheme comprises energy costs based on DA and a dynamic network tariff\footnote[7]{Prices used from \href{https://ssc-fleet.nl/samenwerking/tki-urban-energy/}{FLEET} project's ongoing pilot.} stacked on top of the energy cost component. The dynamic tariff varies based on the current transformer loading as explained in Section \ref{sec:optimisation} and depicted in \figurename{\ref{stacked_tariff}}.

\section{Results}

\begin{figure}[t]
\centering
\includegraphics[width=\linewidth]{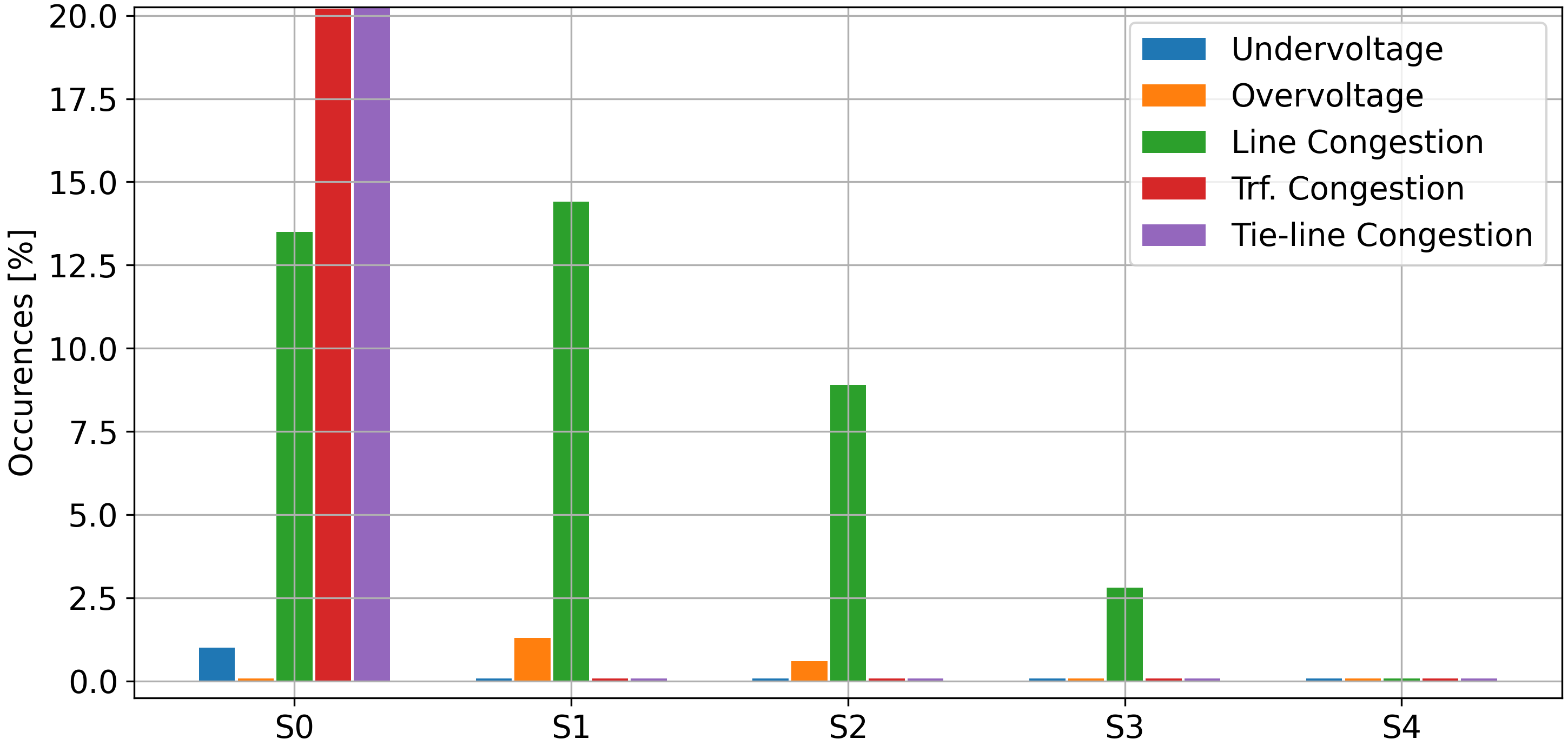}
\caption{Frequency of occurrence for voltage and congestion problems in the modelled LV network for different scenarios .}
\label{fig:S0_4_grid}
\end{figure}

\figurename \ref{fig:S0_4_grid} gives an overview of the frequency of congestion in the lines for the whole simulation period, at the transformer line, transformer, line undervoltage and overvoltage for each scenario. From the uncontrolled scenario S0, it becomes apparent that the amount of congestion is unacceptable with the adopted EV percentage. A minimum capacity of 630 kVA transformer will be needed without smart charging. In S0, out of all total line congestions, 95\% of them occurred in one particular feeder (c.f. \figurename \ref{fig:lv_schematic}). Hence, in S4, the detailed power-flow modelling was used for this particular feeder prone to more congestion, which eventually saved the computational burden needed for solving power-flow for all the feeders.

\begin{figure}[t]
\centerline{\includegraphics[width=\linewidth]{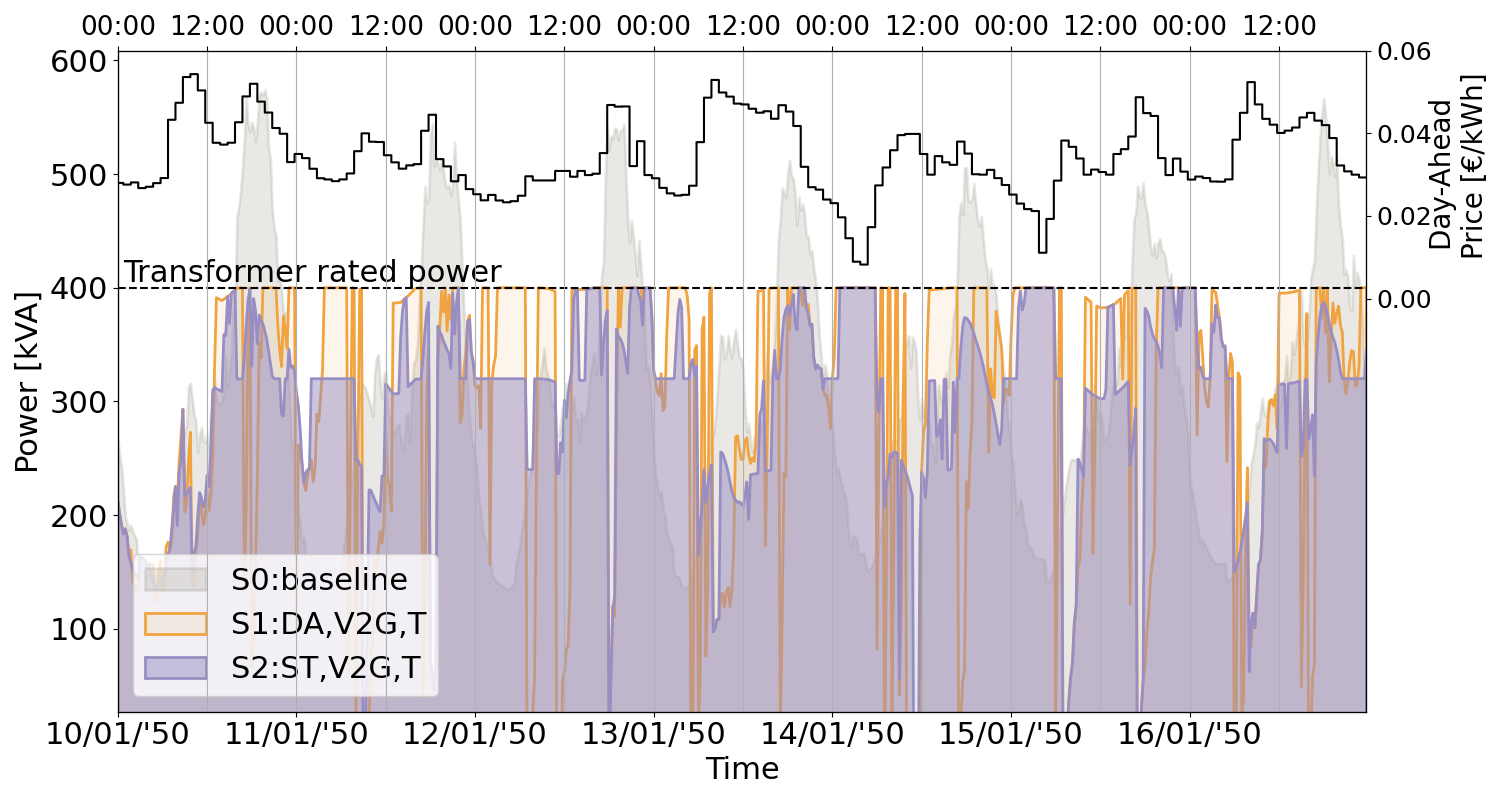}}
\caption{{Aggregated loading comparing uncontrolled (S0) charging with controlled charging using day-ahead (S1) and stacked (S2) tariffs. \textit{Results are obtained from the optimisation model, so power losses are excluded.}}}
\label{fig:S1_trafo}
\end{figure}

In general, scenarios S1-S4 show a clear reduction in the number of grid issues occurrences, signifying the optimisation model's effectiveness. The transformer capacity limit applied to all these scenarios eliminates congestion at the transformer alone but does not cater to localised grid issues down the transformer. Nevertheless, with day-ahead tariffs alone, 6.6\% more congestion is perceived in the lines relative to uncontrolled charging. In contrast, with the stacked tariffs used in S2, a relative reduction of 34.1\% and 38.2\% is perceived compared to S0 and S1, respectively. These results can be partially explained with \figurename \ref{fig:S1_trafo}, which plots the aggregated transformer power for S0, S1 and S2 together with the day-ahead prices and the transformer rated power. It can be deduced that the day-ahead tariff scenario (S1) hits the transformer line more than three times as often compared to its stacked tariff counterpart (S2). This might lead to more frequent congestion in the downstream lines as perceived in the results of \figurename \ref{fig:S0_4_grid}. As a result, it stresses the need for the modelled transformer limit when dynamic electricity tariffs are applied.

The dynamic network component in the stacked tariff reduces that dependency to some extent by stimulating valley-filling more. The tariff effectiveness also applies to mono-directional (V1G) charging as congestion further reduces to 2.8\%. In this study, little attention was devoted to optimising the different price levels, such that the positive effect of the stacked tariff could potentially be further increased. Nevertheless, even with stacked tariffs, high shares of V2G can increase the need for enforcing grid constraints to areas prone to congestion, which was used in S4. The remaining 5.5\% of congestion occurred in the feeders where detailed power flow was not modelled; hence no constraints enforced the networks' bus voltage and line's current ratings. 

The question remains how these grid constraints, V2G, and the stacked tariffs influence the interests of the DSO along with CPO profits and EV user satisfaction. Tab. \ref{tab:interests} gives a clear overview of these benefits, including the power losses, root means square (RMS) transformer loading, and CPO energy tariff costs, all relative to the uncontrolled charging scenario. The last column of Tab. \ref{tab:interests} shows the collective EV user satisfaction, which is affected by the percentage of vehicles charged to the maximum value of SOC during departure. The darker (greener) the cells of Tab. \ref{tab:interests} are, the more beneficial the results are to the corresponding stakeholder. 
From the DSO perspective, the objective function intends to reduce power losses and stimulate valley-filling by applying a stacked tariff mechanism. A lower RMS loading would, in principle, imply more balanced transformer loading as extreme high loading, positive or negative (in the case of V2G), is squared in the RMS calculation. As the square of the current values determines power losses, the power losses in Tab. \ref{tab:interests} follow a similar trend. 
Compared to day-ahead tariffs (S1), the stacked tariff mechanism reduces power losses in all three cases (S2, S3, S4). However, for S2, the power losses are still higher than the baseline case (S0). It becomes apparent that mono-directional charging, as in S3 and power flow modelling, as in S4, have a beneficial effect on the amount of power losses. Nevertheless, the effect remains marginal compared to uncontrolled charging, as a reduction of around 0.5\% is measured. The second column in Tab. \ref{tab:interests} indicates a strong reduction of the RMS loading in the case of stacked tariffs (S2-S4), implying more valley-filling.

From the CPO perspective, Tab. \ref{tab:interests} addresses three things. First, smart charging decreases the costs by more than 40\% in the case of V2G. With V1G only, less capacity can be moved to cheaper times, as no discharging takes place, leading to higher costs. Second, with stacked tariff pricing, higher costs are obtained concerning day-ahead tariff pricing as some loading on most price-favourable moments is shifted to other times of the day. Based on the electricity component, this implies a relative increase of 9.3\% cost between S1 and S2.
Nevertheless, the gains caused by reducing the network component costs of the stacked tariff are not analysed. They could imply higher benefits to the CPO due to its dynamic nature compared to day-ahead tariffs with fixed network costs. Third, adding additional grid constraints to the model does not lower profits significantly, as the difference between S2 and S4 is only 0.04\%. 

From Tab. \ref{tab:interests}, it becomes clear that with V2G, more than 96\% of all charging transactions could be classified as successful. The non-successful transactions have a SOC lower than what could potentially be reached. Compared to V2G, V1G scores worse, as 6.67\% of all charging sessions fail to obtain their full charging potential. This happens mostly around 6 pm, as no V2G is available to create peak-reduction and thus charging capacity for EVs connected for a short time. Consequently, some cars will need to delay charging, resulting in undercharged batteries if the connection time and, thus, flexibility provision is too little.

\begin{table}[t]
\caption{Stakeholder benefits related to each scenario relative to baseline scenario S0.} 
\vspace*{ - 8 mm}
\begin{center}
\begin{tabular}{crrrr}
\hline
{\color[HTML]{3F3F3F} \textbf{\begin{tabular}[c]{@{}c@{}}Scen. \\ Label\end{tabular}}} &
  \multicolumn{1}{c}{{\color[HTML]{3F3F3F} \textbf{\begin{tabular}[c]{@{}c@{}}Power \\ Loss {[}rel\%{]}\end{tabular}}}} &
  \multicolumn{1}{c}{{\color[HTML]{3F3F3F} \textbf{\begin{tabular}[c]{@{}c@{}}RMS Trafo \\ Load {[}rel\%{]}\end{tabular}}}} &
  \multicolumn{1}{c}{{\color[HTML]{3F3F3F} \textbf{\begin{tabular}[c]{@{}c@{}}CPO  \\ Costs {[}rel\%{]}\end{tabular}}}} &
  \multicolumn{1}{c}{{\color[HTML]{3F3F3F} \textbf{\begin{tabular}[c]{@{}c@{}}EV Full\\ SOC {[}abs\%{]}\end{tabular}}}} \\ \hline\hline
{\color[HTML]{3F3F3F} \textbf{S1}} &
  \cellcolor[HTML]{FFEF9C}{\color[HTML]{3F3F3F} +4.83} &
  \cellcolor[HTML]{FFEF9C}{\color[HTML]{3F3F3F} -1.81} &
  \cellcolor[HTML]{63BE7B}{\color[HTML]{3F3F3F} -48.03} &
  \cellcolor[HTML]{B4D88C}{\color[HTML]{3F3F3F} 96.19} \\ 
{\color[HTML]{3F3F3F} \textbf{S2}} &
  \cellcolor[HTML]{A2D188}{\color[HTML]{3F3F3F} +1.67} &
  \cellcolor[HTML]{7DC680}{\color[HTML]{3F3F3F} -5.50} &
  \cellcolor[HTML]{8BCB84}{\color[HTML]{3F3F3F} -43.34} &
  \cellcolor[HTML]{63BE7B}{\color[HTML]{3F3F3F} 99.21} \\ 
{\color[HTML]{3F3F3F} \textbf{S3}} &
  \cellcolor[HTML]{63BE7B}{\color[HTML]{3F3F3F} -0.48} &
  \cellcolor[HTML]{63BE7B}{\color[HTML]{3F3F3F} -6.24} &
  \cellcolor[HTML]{FFEF9C}{\color[HTML]{3F3F3F} -29.45} &
  \cellcolor[HTML]{FFEF9C}{\color[HTML]{3F3F3F} 93.33} \\ 
{\color[HTML]{3F3F3F} \textbf{S4}} &
  \cellcolor[HTML]{63BE7B}{\color[HTML]{3F3F3F} -0.49} &
  \cellcolor[HTML]{73C37E}{\color[HTML]{3F3F3F} -5.78} &
  \cellcolor[HTML]{8BCB84}{\color[HTML]{3F3F3F} -43.30} &
  \cellcolor[HTML]{6CC17D}{\color[HTML]{3F3F3F} 98.89} \\ \hline
\end{tabular}
\label{tab:interests}
\end{center}
\end{table}

 \vspace{-.15 cm}

\section{Conclusion}

This study assessed the application of dynamic electricity and network tariffs, primarily concerning the DSO next to the EV owner and CPO. The exact results of this study are case-study specific but serve as a proof-of-concept such that some key takeaways can be drawn.
It became apparent that the dynamic stacked tariff can have multiple benefits for the DSO while reducing CPO costs compared to uncontrolled charging (to over 40\%) and maintaining acceptable battery SOC levels (more than 98\% satisfaction). The stacked tariffs tend to decrease power losses and reduce congestion problems even downstream the lines, such that power flow modelling becomes less relevant. The latter could still be applied to areas of the grid that are more prone to voltage and congestion without violating other stakeholder interests. These analyses also clarified the (dis)advantages of V2G. V2G facilitates more EVs in power networks due to peak reduction at critical moments. Besides, CPO costs can decrease by more than 20\% using V2G compared to V1G. 
Nevertheless, it might increase power losses slightly and increases local line congestion. Using V2G only during peak moments to allow the charging of EVs with less flexibility would be a better trade-off to make than using V2G to maximise CPO profits. {Future research could regard this implementation and include more uncertainty factors (e.g. SOC, $t^{\text{dep}}$). Fairness and transparency issues related to dynamic tariffs should be addressed as well.} This study concludes by stating that with dynamic tariffs, V2G and locally-applied power flow modelling, it is possible to optimally use EVs' charging flexibility and significantly reduce grid reinforcement towards 2050.

 \vspace{-.1 cm}

\section*{Acknowledgment}

This work is completed under the scope of the \href{https://tki-robust.nl/}{ROBUST} project, which is funded by the MOOI 2020 Top Sector Energy subsidy programme by the Ministry of Economic Affairs and Climate Policy, executed by the Netherlands Enterprise Agency.

\vspace{-0.3 cm}
\bibliographystyle{IEEEtran}

\bibliography{references.bib}

\end{document}